\documentclass[12pt,preprint]{aastex}
\usepackage{natbib}

\begin{document}

\title{The Globular Cluster Mass Function as a Remnant of Violent Birth}

\author{Bruce G. Elmegreen}
\affil{IBM T. J. Watson Research Center, 1101 Kitchawan Road, Yorktown
Heights, New York 10598 USA} \email{bge@us.ibm.com}

\begin{abstract}
The log-normal shape of the mass function for metal-poor halo globular
clusters is proposed to result from an initial $M^{-2}$ power law
modified rapidly by evaporation, collisions with clouds, and mutual
cluster interactions in the dense environment of a redshift $z\sim5-15$
disk galaxy. Galaxy interactions subsequently spray these clusters into
the galaxy group environment, where they fall into other growing
galaxies and populate their halos. Clusters forming later in $z\sim2-5$
galaxies, and those formed during major mergers, produce metal-rich
globulars. Monte Carlo models of evolving cluster populations
demonstrate the early formation of a log-normal mass function for
typical conditions in high-redshift galaxies.
\end{abstract}

\section{Introduction}
The globular cluster mass function (GCMF) in present-day galaxy halos
is approximately log-normal with a peak $M_{\rm
p}\sim10^{5.3}\;M_\odot$ \citep[see reviews in][]{mc03,bs06}. The
origin of this peaked distribution is not understood. Wherever massive
dense clusters like globular clusters (GCs) are formed today, they have
a power-law mass function like $dN/dM\propto M^{-2}$, or a Schechter
function with a similar power law at low mass and a cutoff at high mass
\citep[see review in][]{gieles09}. Consequently, one theory for halo
GCs is that they begin with a power-law mass function and then lose
their low-mass members through dispersal over a Hubble time
\citep{fr77, ot95,ee97,fz01}. The dispersal rate works out about right
if the disruption process is thermal evaporation \citep{mf08}.

The problem with this model is that the GCMF does not vary with radius
in several nearby galaxies \citep{tam06,jor07}, and evaporation is
expected to occur faster in the inner regions where tidal forces are
larger, thereby shifting the peak toward higher masses there. A model
in which GCs have a small range of pericenters can fix this problem
\citep{fz01}, but the resulting GC velocities disagree with
observations in M87 \citep{ves03b}.

What is expected to drive the radial gradient in evaporation rate is a
gradient in the tidal density \citep{gb08}, which is the average GC
density inside the GC tidal radius. This tidal density is difficult to
observe directly and is not necessarily proportional to the average
density inside the half-light radius, which is observed directly. The
half-light density does not correlate well with radial distance. Noting
this, \citet{chan07} and \citet{mf08} fit the GCMFs in M104 and the
Milky Way to evolved Schechter functions for three bins of half-light
density using GC evaporation rates proportional to the square roots of
these densities. The results are consistent with faster evaporation,
i.e., higher peak mass, at greater half-light density, regardless of
galactocentric radius. \citet{mf08} also fit the Milky Way GCMF for
three bins of tidal density using evaporation rates given by that
density.

An alternative model is that GCs are born with a peaked mass function
and then keep it over a Hubble time \citep{ves03b,pg05}. Log-normal
mass functions evolve somewhat self-similarly during evaporation, and
their peak mass and width may even converge to the observed values
\citep{ves98}. The peak mass would also be uniform with galactocentric
radius after a while. Initially peaked GCMFs could result from a lack
of low-mass clouds in the early GC environment \citep{pg07}. Another
model considers variable star formation efficiencies with a greater
probability for lower cluster masses to disperse when the gas leaves
\citep{parm08,baum08}. A third model is that low mass clusters were
born with lower central concentrations and so evaporated more quickly
than high mass clusters \citep{vz03}. There is no direct evidence for
any of these models because all known clusters today are born with
power-law mass functions. For one of these models to be viable, the
qualitative nature of cluster formation would have to be different in
the early Universe.

We consider here a model where halo GCs form with power-law mass
functions that are quickly converted into peaked functions in very
dense cloudy environments. The GCMF stays peaked and insensitive to
environment thereafter \citep{ves98}. This model involves the same
physical process of cluster formation that is present today, i.e.,
gravitational collapse in giant gas complexes, and the cluster-forming
cloud-cores are probably similar as well, considering that GC
densities, masses, and IMFs are not unusual. What differed in young
galaxies was a much denser and more turbulent interstellar medium (ISM)
than we have in main galaxy disks today \citep{fs06,g06,law07}. GC
formation in $z>10$ dwarf galaxies was discussed by \citet{br04};
bulge-GC formation in $z\sim2$ galaxy clumps was discussed by
\citet{shap10}, and simulations of GC formation in young galaxy disks
were made by \citet{krav05}. The main point here is that when the
density is high and the motions are fast, clusters should frequently
collide with cloud clumps and other clusters during their first several
hundred Myrs. These collisions destroy the lowest mass clusters and
produce a log-normal GCMF.

\section{Collisional Dispersal of Clusters in High Density Environments}

Clusters heat up or disperse completely when their potential energy
changes quickly. This can result from rapid gas expulsion or from
collisions with dense clouds and other clusters. In a star-forming
environment, there are many opportunities for collisions. The cluster
dispersal rate therefore starts high after birth, and it decreases as
the star-forming cloud dissipates and all the clusters and dense cores
drift away. We have studied time-dependent cluster collisions elsewhere
\citep{eh10}. Here we consider a time-average collision rate and ask
what it has to be in order to convert a power-law initial cluster mass
function into a peaked mass function during the starburst phase of a
young galaxy.

The collisional disruption rate is taken from \citet{gieles06}, who use
a cluster evolution equation $dM/dt=-M/t_{\rm dis}$ with a disruption
time
\begin{equation}
t_{\rm dis}=37{M^{0.61}\over{\Sigma_{\rm n}\rho_{\rm n}}}\; {\rm
Myr};\label{dis}
\end{equation}
$\Sigma_{\rm n}$ is the average column density of a collision partner,
in $M_\odot\;{\rm pc}^{-2}$, $\rho_{\rm n}$ is the average density of
collision partners in the neighborhood, in $M_\odot\;{\rm pc}^{-3}$,
and $M$ is the cluster mass. Gieles et al. use $\Sigma_{\rm
n}=170M_\odot\;{\rm pc}^{-2}$ for typical GMCs and $\rho_{\rm
n}=0.03M_\odot\;{\rm pc}^{-3}$ for the ambient ISM. Then $\Sigma_{\rm
n}\rho_{\rm n}\sim5.1M_\odot^2\;{\rm pc}^{-5}$ and $t_{\rm
dis}\sim7.3M^{0.61}$ Myr, which is $8.2$ Gyr for $M=10^5\;M_\odot$.

The disruption time is less at higher density. If we require disruption
of $M<10^5\;M_\odot$ clusters within the time span of the gas-rich
phase of a young galaxy disk, which may be 500 Myr
\citep[e.g.,][]{tac08}, then $t_{\rm dis}=500$ Myr and $\Sigma_{\rm
n}\rho_{\rm n}\sim80 M_\odot^2\;{\rm pc}^{-5}$ -- a factor of 16 larger
than the local ISM value used by Gieles et al.. This factor may be
accounted for by a higher average density in the cluster environment
and a higher average column density for likely collision partners. For
example, $\rho_{\rm n}$ could be $\sim10$ times higher in a galaxy with
50\% of the disk mass in the form of clumpy gas \citep{tac10}, and
$\Sigma_{\rm n}$ could be $\sim10^3\; M_\odot$ pc$^{-2}$ or more for
massive molecular cores and other clusters.  If we require that $t_{\rm
dis}$ equals 10 times the dynamical time of the ISM, which is
$10/\left(G\rho_{\rm n}\right)^{0.5}$, then $\Sigma_{\rm n}\rho_{\rm
n}^{0.5}$ has to equal $280\;M_\odot^{1.5}\;{\rm pc}^{-3.5}$ for
$M=10^5\;M_\odot$. Even this is reasonable for an extremely gas-rich
and dense disk where $\rho_{\rm n}$ might be $\sim1\;M_\odot$ pc$^{-3}$
($\sim30$ atoms cm$^{-3}$) in regions one kpc in size.  The clumpy
galaxies observed at $z\sim2$ are a good example of such an environment
\citep{e09}. The ISM density should be even larger at $z\sim5-15$,
where many halo GCs are likely to have formed. Young galaxies are
denser than today's galaxies because the Universe was denser, by the
factor $(1+z)^3\sim1000$ for the redshift of interest here.  Young
galaxies are also smaller than today's galaxies \citep{oe10} and either
represent the inner (dense) parts of today's galaxies or separate
objects that merged into dense spheroids over time.

\subsection{Monte Carlo Simulations with Constant Cluster Birthrate}

The time evolution of cluster populations was determined for Monte
Carlo simulations with constant cluster birthrates and cluster
disruption rates given by
\begin{equation}dM/dt= -M/t_{\rm dis} \;\;;\;\;t_{\rm dis}=\xi M^\gamma;
\label{eq:2}\end{equation} $\xi=37/\left( \Sigma_{\rm n}\rho_{\rm
n}\right)$ could be $\sim0.01$ or smaller for units of column density
in $M_\odot$ pc$^{-2}$ and density in $M_\odot$ pc$^{-3}$. In the
\citet{gieles06} model, $\gamma\sim0.6$; we take $\gamma=0.62$ to be
consistent with the evaporation model in \citet{lamers05}. We also ran
simulations with $\gamma=1$, which is appropriate for \citet{spit87}
evaporation and for \citet{spit58} collisional disruption with a weak
mass-radius relation, as observed by \citet{bastian05} and others. That
is, \citet{gieles06} and \citet{spit58} both derived $t_{\rm
dis}\propto \rho_{\rm cl}/\left(\Sigma_{\rm n}\rho_{\rm n}\right)$ for
collisional disruption with cluster internal density $\rho_{\rm cl}$.
Bastian et al. used $\rho_{\rm cl}\propto M^{0.61}$, but it is also
possible that $\rho_{\rm cl}\propto M$. Both collisional disruption and
cluster evaporation should be rapid in the high density environments of
young GCs. Collisional disruption probably dominates evaporation in the
disk environment \citep{gieles06}.

Values of $\xi$ that give a mass distribution peak $M_{\rm p}$ when the
cluster population has an age $T$ may be determined from the relation
$\xi M_{\rm p}^\gamma=T$. For $T=1000$ Myr and $M_{\rm
p}=10^5\;M_\odot$, $\xi=0.01$ when $\gamma=1$, and $\xi=0.79$ when
$\gamma=0.62$. These $\xi$ are reasonable for starburst conditions in
young, dense galaxies.

\begin{figure}
\epsscale{.55} \plotone{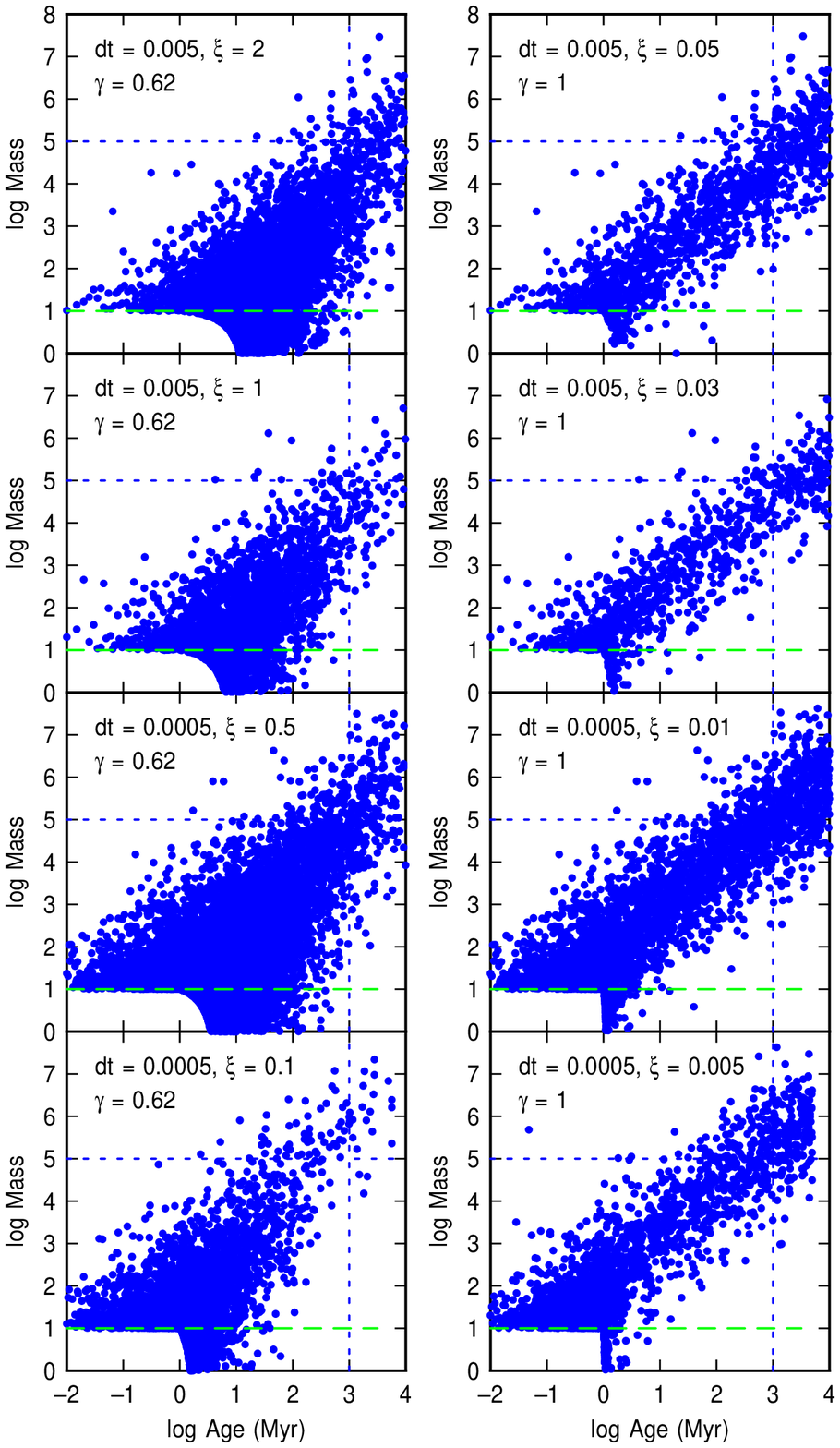} \caption{Distribution of cluster masses
$M$ and ages $T$ after $10^4$ Myr of evolution with continuous cluster
formation in an $M^{-2}dM$ mass function and continuous disruption or
evaporation according to equation (2) with values of $\gamma$ and $\xi$
indicated. The star formation rate is $1/dt$ for model time step $dt$
in Myr. The horizontal dashed line is the lower limit to the mass of a
formed cluster; clusters get lower mass over time because of the
assumed disruption. The blue dotted lines indicate fiducial markers for
age and mass where we expect a GC population should be after the
starburst phase of a young galaxy. The clusters are presumed to be
scattered after $\sim1$ Gyr and then they evolve mostly by slow
evaporation in the halos and bulges of these and other galaxies.
\label{gcform_mtplot7c_both}}
\end{figure}

In the simulations, clusters masses were randomly chosen from an
initial $dn/dM\propto M^{-2}$ mass function that extends from a minimum
mass $M_{\rm min}=10\;M_\odot$ to a maximum mass of $10^8\;M_\odot$.
The cluster formation rate is one cluster per time step, $dt$, measured
in Myr.  At each time step, equation (\ref{eq:2}) is applied to every
cluster, and every cluster mass is reduced accordingly. Clusters with
masses dropping below $0.1M_{\rm min}$ are not followed. The results
after 10 Gyr are shown on a plot of $\log M$ versus $\log T$ for each
current cluster mass $M$ and age $T$ (Fig. \ref{gcform_mtplot7c_both}).
The age is defined to be the current time in the simulation minus the
time of formation of the cluster. We are interested in the distribution
of cluster mass at an age of $\sim500$ Myr or so, when the starburst
ends and the clusters begin to scatter. This distribution can be read
off the plot at $T=500$ Myr. The plot extends to $T=10$ Gyr for
comparison.

Figure \ref{gcform_mtplot7c_both} shows $\log M - \log T$ distributions
of cluster populations for various values of $\xi$ and for
$\gamma=0.62$ on the left and $\gamma=1$ on the right. All of the
distributions have about the same form with more or less attrition over
age in cases with low or high $\xi$, respectively. The upper envelope
of the cluster mass comes from the size-of-sample effect, which states
that the most massive cluster in an $dn/dM\sim M^{-2}$ distribution is
directly proportional to the number of clusters in the sample. This
maximum increases linearly with $T$ on a plot like this because the
time interval $\Delta T$ increases linearly with $T$ for equal
intervals in $\log T$.  The lower envelope of the distributions
increase as $M\propto T^{1/\gamma}$ because that is the mass at which
$t_{\rm dis}=T$.  The dashed green line is $M_{\rm min}=10\;M_\odot$.
For $\gamma=1$, the upper and lower envelopes are parallel so the mass
range of the mass function at any one age remains about constant. For
$\gamma=0.62$, the bottom envelope increases faster than the top
envelope. Two cluster formation rates ($1/dt$) are used to give good
numbers of points at different $\xi$.

Lower $\xi$ makes the characteristic mass of the clusters higher for a
given age. Figure \ref{gcform_mtplot7c_both} has vertical and
horizontal dotted lines at fiducial markers $T=1000$ Myr and
$M=10^5\;M_\odot$. At $T=1000$ Myr with $\gamma=0.62$, $\xi=0.1$ has a
mass distribution centered slightly higher than $M\sim10^5\;M_\odot$,
while $\xi=1$ has a mass distribution centered slightly lower than
this. The value $\xi=0.5$ gives a mass at the peak of the function
$M_{\rm p}\sim10^5\;M_\odot$, as predicted approximately above. For
$\gamma=1$, a value of $\xi\sim0.01$ gives this $M_{\rm p}$ at $T=1000$
Myr, as also predicted.

\begin{figure}
\epsscale{1.0} \plotone{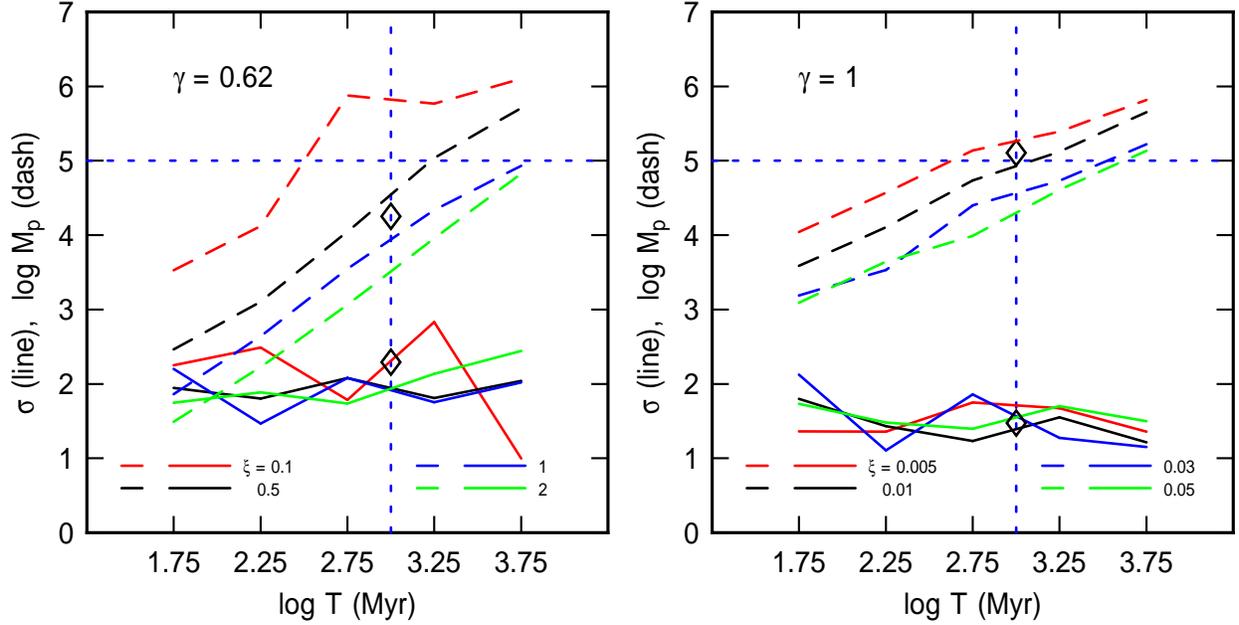} \caption{Fits for peak masses $M_{\rm
p}$ and dispersions $\sigma$ of log-normal cluster mass functions for
0.5 log-age intervals.  Different curves are for different $\xi$, which
is the coefficient in equation (2) for the dispersal time. The peak
mass increases with age because of preferential dispersal of low mass
clusters. Strong cluster disruption, corresponding to small values of
$\xi$, can produce a peaked GCMF with $M_{\rm p}\sim10^5\;M_\odot$
after $\sim1$ Gyr, as indicated by the dotted lines. Diamonds give
$M_{\rm p}$ and $\sigma$ values for the models shown as red crosses in
Fig. 3. \label{gcform_fits}}
\end{figure}

Figure \ref{gcform_fits} shows log-normal fits to the cluster mass
functions versus cluster age in four age intervals, $\log T=2-2.5$,
$2.5-3$, $3-3.5$, and $3.5-4$ for $T$ in Myr. The trend of increasing
mean $\log M$ with increasing age (Figure \ref{gcform_mtplot7c_both})
is evident in Figure \ref{gcform_fits}. The scatter in the diagram is
from the small numbers of clusters in the age intervals (50 to 500,
depending on $\xi$ and $\gamma$).  The solutions give the desired
$M_{\rm p}\sim10^5$ when $T\sim10^3$ Myr if $\xi$ is small.

\subsection{Monte Carlo Simulations with Cluster Birth for 500 Myr}

\begin{figure}
\epsscale{1.0} \plotone{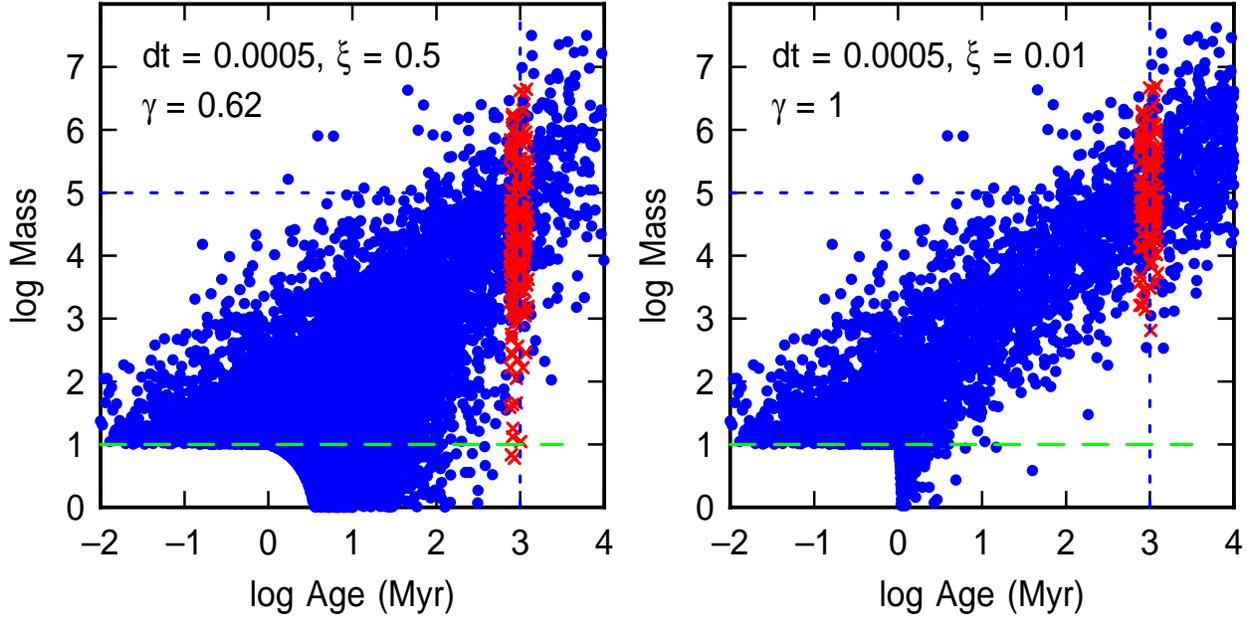} \caption{Masses and ages of cluster
populations after $10^4$ Myr with continuous cluster formation and
dispersal (blue dots, as in Figure 1). Surviving clusters (red crosses)
for a second model in which cluster formation and full-rate disruption
end after 500 Myr, and then cluster disruption diminishes in proportion
to the total mass of the remaining clusters. This second model is
viewed after $10^3$ Myr. It represents a case where clusters are formed
and disrupted during a Gyr time span in a small dense galaxy at high
redshift.  The clusters end this phase with a log-normal mass function
and presumably scatter into other forming galaxies over time,
preserving this mass function. \label{gcform_mtplot7c_stop}}
\end{figure}

\begin{figure}
\epsscale{1.0} \plotone{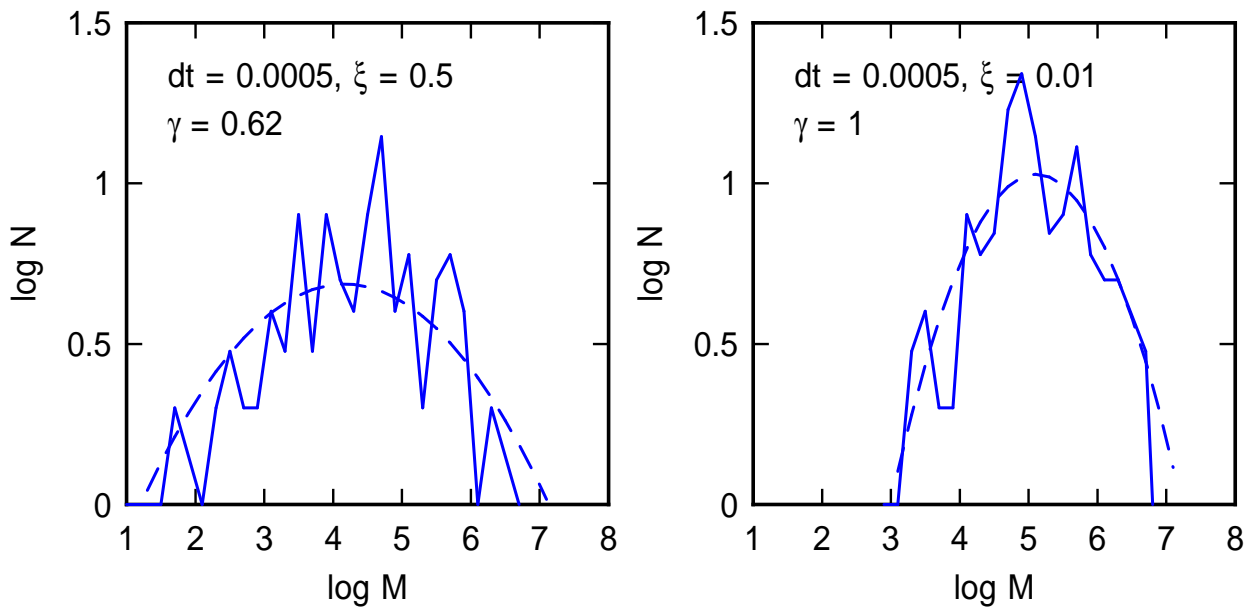} \caption{Mass distribution functions of
the red crosses shown in Figure 3, along with log-normal fits. The
$\gamma=0.62$ case has a tail of low mass clusters for the assumed
epoch (1 Gyr) and disruption rate ($\xi$), but both are reasonably
peaked at around $M_{\rm p}\sim10^5\;M_\odot$, making this process
viable as a mechanism to inject GCs with this ``initial'' mass function
into the converging flows of the early universe. \label{gcform_gcmf}}
\end{figure}

Figure \ref{gcform_mtplot7c_stop} shows another case. The blue dots
come from previous models (Figure \ref{gcform_mtplot7c_both}). The red
crosses are for a model where cluster formation occurs for the first
500 Myr and cluster disruption uses equation (\ref{eq:2}), but then
star formation stops and the disruption rate decreases in proportion to
the remaining cluster mass as
\begin{equation}dM/dt=-\left({{M_{\rm tot}(t>500\;{\rm Myr})}
\over{M_{\rm tot}(t=500\;{\rm Myr})}}\right){{M}\over{t_{\rm
dis}}}\end{equation} for times between 500 Myr and 1000 Myr. This model
simulates normal cluster formation and disruption during 500 Myr when
dense gas is present, followed by no cluster formation and diminished
disruption for another 500 Myr after the gas disperses. Only the
log-normal GCMF remains; fits are shown in Figure \ref{gcform_gcmf}.
Figure \ref{gcform_fits} shows black symbols for the peak masses and
dispersions in this second model at $T=2.75-3.25$ Myr; they should be
compared with the black curves, which have the same $\gamma$ and $\xi$.

\section{Discussion}

We propose that old halo GCs formed by normal processes much like
clusters form today, but that all of these processes occurred in young
galaxies at very high densities, higher than today's ISM densities by
factors of 10 to 100. Then collisional disruption and evaporation was
rapid enough to produce a peaked mass function in the young galaxy.
Subsequent evolution by evaporation in lower density environments
preserved this function, although slight changes in peak mass and width
probably occurred.

Given this basic model, there are many possibilities for the delivery
of these clusters into modern galaxies.  The dense galaxies they formed
in are most likely not the same as the galaxies or inner parts of the
galaxies that currently host them. Their presence in galaxy halos today
implies that they were delivered to the host in a non-dissipative way,
perhaps by infall along with most of the host's other baryons.  For
example, the oldest GCs could have formed in small dense galaxies that
formed as condensations in larger-scale cold gas flows. The cold flows
make today's spiral disks \citep{dekel09,atm09,keres09}, and the GCs
along with other condensations in the flow populate modern halos. The
small dense galaxies could also have collided before they entered the
modern galaxy's potential well, freeing up the GCs which would then
enter the well as free-floaters along with the other material. Some
small galaxies are still adding GCs to modern halos
\citep[e.g.,][]{carraro07,gao07,casetti09,smith09}. The oldest GC
populations in elliptical galaxies presumably formed in small dense
galaxies and fell into spiral halos too in the same way, but then ended
up in ellipticals after major mergers of the spirals. These oldest GCs
would be the blue, metal-poor populations in spiral and elliptical
halos.  GCs formed during major mergers would be redder and more
metal-rich \citep{bs06}.

The uniform properties of old halo globular clusters (color, density,
IMF, peak GCMF mass, metallicity) follow from this model if we
postulate that GCs are the first examples of star formation at high
enough metallicity to make a normal IMF. Then stellar evolution will
not remove excessive amounts of gas and cause the cluster to come
unbound. Top-heavy IMFs, such as those thought to produce the
carbon-enhanced metal-poor stars \citep{tumlinson07,komiya09}, lose too
high a fraction of their mass during stellar evolution to keep a
cluster bound. Boundedness would require them to occupy the nuclei of
small galaxies so they can retain their stellar wind material in the
galactic potential well. Then contamination from subsequent generations
of stars would also enrich them \citep[e.g.,][]{marc07,bailin09}. For
clusters born in small dense disks, the uniformity of both the clusters
and the GCMFs suggest that the first epoch of normal IMFs occurred when
galaxies were still small and dense, and this was long before today's
spirals were assembled.

The bulge GCs of modern spirals could have a similar origin as the halo
GCs, but there is no similar constraint on the constancy of their GCMFs
over a range of galactocentric radii. Also, the bulge GCs formed in
metal-rich environments and they are now located deep in the potential
wells of their current hosts. Thus they need not have formed in
separate small galaxies. They could have formed in their hosts and
settled to the center after energy dissipation. This is the model by
\citet{shap10}. They proposed that bulge GCs formed in the dense clumps
of young massive $z\sim2$ galaxies, much like we propose that
metal-poor GCs formed in the dense clumps of young and small
$z\sim5-15$ galaxies. The local environment is about the same for each,
namely high-density clumpy gas, so collisional disruption would have
occurred quickly for the Shapiro et al. model too. The densities in the
massive clumps of $z\sim2$ spirals are not expected to have been as
high as the densities in the disk clumps of smaller galaxies at higher
redshift, however. Thus collisional disruption of low mass GCs may not
have been as rapid for the Shapiro et al. model.

Our model accounts for the number of halo GCs today if a high fraction
of early star formation made bound clusters. Such a high fraction is
expected when the ISM pressure and velocity dispersion are large
\citep{e08}. The space density of GCs today is $\sim8$ Mpc$^{-3}$
\citep{pm00}. There could have been a factor of $\sim2$ more when they
formed, considering evaporation \citep{ves98}, but blue halo GCs are
only half of the total. The co-moving space density of clumpy galaxies
at $z\sim1-2$ is $n_{\rm g}\sim2\times10^{-3}$ Mpc$^{-3}$ \citep{e07}.
Each galaxy has $\sim5$ clumps of stellar mass $\sim10^8\;M_\odot$,
which would be $N_{\rm c}\sim5\times10^3$ clusters of mass
$\sim10^5\;M_\odot$ each. Thus the space density of massive clusters
would have been $n_{\rm g}N_{\rm c}\sim10$ Mpc$^{-3}$, with a
factor-of-3 uncertainty either way. \citet*{shap10} got the same result
a different way. If metal-poor GCs formed in younger, smaller galaxies,
then the $M_{\rm gal}^{-2}$ cosmological galaxy mass function means
that each log interval of galaxy mass contains the same total mass.
Thus we would get these same $\sim10$ GC per Mpc$^3$ for GC formation
in galaxies 10 or 100 times less massive at higher redshifts.

Modern examples of this model may occur in dense galactic nuclear
regions where we predict a rapid destruction of low mass clusters
because of heightened collision and evaporation rates.

Acknowledgements: I am grateful to Enrico Vesperini and Tanuka
Chattopadhyay for discussions about the GCMF, and to Dean McLaughlin
for comments on the manuscript.

\clearpage

\begin{thebibliography}{}

\bibitem[Agertz, Teyssier \& Moore(2009)]{atm09} Agertz, O., Teyssier, R., Moore, B. 2009, MNRAS,
397, L64

\bibitem[Bailin \& Harris(2009)]{bailin09} Bailin, J., \& Harris, W.E. 2009, ApJ, 695, 1082

\bibitem[Bastian et al.(2005)]{bastian05} Bastian, N., Gieles, M., Lamers, H. J. G. L. M., Scheepmaker, R. A.,
\& de Grijs, R. 2005, A\&A, 431, 905

\bibitem[Baumgardt, Kroupa \& Parmentier(2008)]{baum08} Baumgardt, H., Kroupa, P., \& Parmentier, G. 2008, MNRAS,
384, 1231

\bibitem[Brodie \& Strader(2006)]{bs06} Brodie, J.P., \& Strader, J. 2006, ARA\&A, 44, 193

\bibitem[Bromm(2004)]{br04} Bromm, V. 2004, ASPC, 322, 499

\bibitem[Carraro et al. (2007)Carraro, Zinn \& Moni Bidin]{carraro07} Carraro, G., Zinn, R., \& Moni Bidin, C. 2007,
\textit{A\&A}, 466, 181

\bibitem[Casetti-Dinescu(2009)]{casetti09} Casetti-Dinescu, D.I. et al. 2009, ApJ, 701, L29

\bibitem[Chandar, Fall \& McLaughlin(2007)]{chan07} Chandar, R., Fall, S. M., \& McLaughlin, D.E. 2007, ApJ,
668, L119

\bibitem[Dekel et al.(2009)Dekel, Sari \& Cervino]{dekel09} Dekel, A., Sari, R., \& Ceverino, D.\ 2009, ApJ, 703, 785

\bibitem[Elmegreen(2008)]{e08} Elmegreen, B.G. 2008, ApJ, 672, 1006

\bibitem[Elmegreen \& Efremov(1997)]{ee97} Elmegreen, B.G., \& Efremov, Y.N. 1997, ApJ, 480, 235

\bibitem[Elmegreen et al.(2007)]{e07} Elmegreen, B.G., Elmegreen, D.M. Ravindranath, S., Coe,
D.A. 2007, ApJ, 658, 763

\bibitem[Elmegreen et al.(2009)]{e09} Elmegreen, B.G., Elmegreen, D.M., Fernandez, M.X., \&
Lemonias, J.J., 2009, ApJ, 692, 12

\bibitem[Elmegreen \& Hunter(2010)]{eh10} Elmegreen, B.G., \& Hunter, D.A. 2010, ApJ, 712, 604

\bibitem[Fall \& Rees(1977)]{fr77} Fall, S.M., \& Rees, M.J. 1977,  MNRAS, 181, 37

\bibitem[Fall \& Zhang(2001)]{fz01} Fall, S.M., \& Zhang, Q. 2001, ApJ, 561, 751

\bibitem[F\"orster Schreiber et al.(2006)]{fs06} F\"orster Schreiber, N. M. et al. 2006,
ApJ, 645, 1062

\bibitem[Gao et al.(2007)]{gao07} Gao, S., Jiang, B.-W., \& Zhao, Y.-H. 2007, \textit{ChJAA},
7, 111

\bibitem[Genzel et al.(2006)]{g06} Genzel, R. et al. 2006, Nature, 442, 786

\bibitem[Gieles(2009)]{gieles09} Gieles, M. 2009, MNRAS, 394, 2113

\bibitem[Gieles et al.(2006)]{gieles06} Gieles, M., Portegies Zwart, S. F., Baumgardt, H.,
Athanassoula, E., Lamers, H. J. G. L. M., Sipior, M., \& Leenaarts, J.
2006, MNRAS, 371, 793

\bibitem[Gieles \& Baumgardt(2008)]{gb08} Gieles, M., \& Baumgardt, H. 2008, MNRAS, 389, L28

\bibitem[Jord\'an et al.(2007)]{jor07} Jordan, A., McLaughlin, D.E., C\^ot\'e, P., Ferrarese, L.,
Peng, E.W., Mei, S., Villegas, D., Merritt, D., Tonry, J.L., \& West,
M.J. 2007, ApJS, 171, 101

\bibitem[Keres et al.(2009)]{keres09} Keres, D., Katz, N., Fardal, M., Dav\'e, R., \& Weinberg,
D.~H.\ 2009, MNRAS, 395, 160

\bibitem[Komiya et al.(2009)]{komiya09} Komiya, Y., Suda, T., Fujimoto, M.Y. 2009, ApJ, 694, 1577

\bibitem[Kravtsov \& Gnedin(2005)]{krav05} Kravtsov, A.V., \& Gnedin, O.Y. 2005, ApJ, 623, 650

\bibitem[Lamers et al.(2005)]{lamers05} Lamers, H.J.G.L.M., Gieles, M., Bastian, N., Baumgardt, H.,
Kharchenko, N.V., \& Portegies Zwart, S. 2005, A\&A, 441, 117

\bibitem[Law et al.(2007)]{law07} Law D. R., Steidel C. C., Erb D. K., Larkin J. E., Pettini
M., Shapley A. E., \& Wright S. A., 2007, ApJ, 669, 929

\bibitem[Marcolini et al.(2007)]{marc07} Marcolini, A., Sollima, A., D'Ercole, A., Gibson, B. K., Ferraro, F. R. 2007, \textit{MNRAS}, 382, 443


\bibitem[McLaughlin(2003)]{mc03} McLaughlin, D.E. 2003, in Extragalactic Globular Cluster
Systems, ESO Astrophysics Symposia, ed. M. Kissler-Patig. Berlin:
Springer-Verlag, p. 329

\bibitem[McLaughlin \& Fall(2008)]{mf08} McLaughlin, D.~E., \& Fall, S.~M. 2008, ApJ, 679, 1272

\bibitem[Okazaki \& Tosa(1995)]{ot95} Okazaki, T., \& Tosa, M. 1995, MNRAS, 274, 48

\bibitem[Oesch et al.(2010)]{oe10} Oesch, P. A., Bouwens, R. J., Carollo, C. M., Illingworth,
G. D., Trenti, M., Stiavelli, M., Magee, D., Labb\'e, I., \& Franx, M.
2010, ApJ, 709, L21

\bibitem[Parmentier \& Gilmore(2005)]{pg05} Parmentier, G., \& Gilmore, G. 2005, MNRAS, 363, 326

\bibitem[Parmentier \& Gilmore(2007)]{pg07} Parmentier, G., \& Gilmore, G. 2007, MNRAS, 377, 352

\bibitem[Parmentier et al.(2008)]{parm08} Parmentier, G., Goodwin, S. P., Kroupa, P., \& Baumgardt,
H. 2008, ApJ, 678, 347

\bibitem[Portegies Zwart \& McMillan(2000)]{pm00} Portegies Zwart, S. F., \& McMillan, S. L. W. 2000, ApJ, 528, L17

\bibitem[Smith et al.(2009)]{smith09} Smith, M. C., Evans, N. W., Belokurov, V., Hewett, P. C.,
Bramich, D. M., Gilmore, G., Irwin, M. J., Vidrih, S., \& Zucker, D. B.
2009, MNRAS, 399, 1223

\bibitem[Spitzer(1958)]{spit58} Spitzer, L., Jr. 1958, ApJ, 127, 17

\bibitem[Spitzer(1987)]{spit87} Spitzer, L., Jr. 1987, Dynamical Evolution of Globular
Clusters (Princeton: Princeton University Press).

\bibitem[Shapiro, Genzel \& F\"orster Schreiber(2010)Shapiro et al.]{shap10}
Shapiro, K.L., Genzel, R., \& F\"orster Schreiber, N.M.
2010, MNRAS, in press

\bibitem[Tacconi et al.(2008)]{tac08} Tacconi, L.J. et al. 2008, ApJ, 680, 246

\bibitem[Tacconi et al.(2010)]{tac10} Tacconi, L.J. et al. 2010, Nature, 463, 781

\bibitem[Tamura et al.(2006)]{tam06} Tamura, N., Sharples, R.M., Arimoto, N., Onodera, M., Ohta,
K., \& Yamada, Y. 2006, MNRAS, 373, 588

\bibitem[Tumlinson(2007)]{tumlinson07} Tumlinson, J. 2007, ApJ, 665, 1361

\bibitem[Vesperini(1998)]{ves98} Vesperini, E. 1998, MNRAS, 299, 1019

\bibitem[Vesperini \& Zepf(2003)]{vz03} Vesperini, E., \& Zepf, S. E. 2003, ApJ, 587, L97

\bibitem[Vesperini et al.(2003)]{ves03b} Vesperini, E., Zepf, S. E., Kundu, A., \& Ashman, K. M.
2003, ApJ, 593, 760

\end{thebibliography}
\end{document}